\documentclass[12pt]{article}
\newcommand{\fnd}[2]{\frac{\textstyle #1}{\textstyle #2}}
\newcommand{\xrm}[1]{{\textstyle \mbox{\rm #1}}}
\newcommand{\bm}[1]{\mbox{\boldmath $#1$}}
\newcommand{\abs}[1]{\left| #1\right|}
\newcommand{\ket}[1]{\mbox{$\left| #1\right\rangle$}}
\newcommand{\bracket}[2]{\mbox{$\left\langle #1\left| #2\right.\right
\rangle$}}
\newcommand{\bra}[1]{\mbox{$\left\langle #1\right|$}}
\newcommand{\x}[1]{{\textstyle #1}}
\begin{document} \baselineskip .7cm
\title{Modified Breit-Wigner formula for mesonic resonances describing
OZI decays of confined $q\bar{q}$ states\\ and\\
the light scalar mesons}
\author{
Eef van Beveren\\
{\normalsize\it Centro de F\'{\i}sica Te\'{o}rica}\\
{\normalsize\it Departamento de F\'{\i}sica, Universidade de Coimbra}\\
{\normalsize\it P-3000 Coimbra, Portugal}\\
{\small eef@teor.fis.uc.pt}\\ [.3cm]
\and
George Rupp\\
{\normalsize\it Centro de F\'{\i}sica das Interac\c{c}\~{o}es Fundamentais}\\
{\normalsize\it Instituto Superior T\'{e}cnico, Edif\'{\i}cio Ci\^{e}ncia}\\
{\normalsize\it P-1049-001 Lisboa Codex, Portugal}\\
{\small george@ajax.ist.utl.pt}\\ [.3cm]
{\small PACS number(s): 11.80.Et, 12.40.Yx, 13.75.Lb, 14.40.-n}\\ [.3cm]
{\small hep-ex/0106077}
}
\date{\today}
\maketitle

\begin{abstract}
A general expression resembling Breit-Wigner formulae
is derived for the description of resonances which appear in meson-meson
scattering.
Starting point is a unitarised meson model, but reduced to a simpler form
and freed from the specific assumption about the confining force.
The parameters of the resulting ``Resonance-Spectrum Expansion'' are directly
related to the confinement spectrum and the mechanism of $^3P_0$
valence-quark-pair creation for OZI-allowed hadronic
decay, and not to the central positions and widths of resonances.
The method also provides a straightforward explanation for the origin of the
light scalar mesons without requiring extra degrees of freedom.
\end{abstract}
\clearpage

\section{Introduction}

Lattice QCD in principle offers the most direct way to link to experiment
what we believe to be the fundamental theory of strong interactions.
However, for the time being only quenched calculations are capable of making,
with reasonable accuracy, predictions for e.g.\ masses of mesons, as a result
of the confinement sector of QCD
\cite{Weingarten94,Weingarten83,Hamber82,Hamber81}.
Relating such predictions to experimental data is quite a controversial issue
though, since the states obtained in quenched lattice QCD (qlQCD) are
manifestly stable, a circumstance which is clearly not the case in
experiment, at least for most mesons.
The reason of course is the difficulty of incorporating quark-pair creation in
qlQCD, thus impeding the process of OZI-allowed strong decay, which is
responsible for the large widths of many mesonic resonances.
But also the central positions of such resonances may be quite displaced due
to the effects of strong decay \cite{BDR83},
when compared with the stable qlQCD states.
Moreover, the contributions of \em virtual \em \/hadronic decays through quark
loops owing to the presence of \em closed \em \/OZI-allowed thresholds, 
which should also lead to real mass shifts, are not fully accounted for in
qlQCD \cite{Ali0105015,Aoki99,Kuramashi99}. 
An additional complication is the observation that not even the \em number \em
\/of $q\bar{q}$ states in the $J^{P}=0^{+}$ sector of the qlQCD predictions
seems to agree with experiment \cite{LW99}, thus contributing to the general
confusion concerning especially the light scalar mesons.

It is evidently a very unsatisfactory state of affairs that, notwithstanding
the ever improving accuracy of the numerical predictions in qlQCD, these
cannot be trusted to unmistakably confirm possible signals of new physics.
In the present paper, we shall propose an alternative method of data analysis,
which does allow for an accurate link between qlQCD, or any other,
model-dependent formulation of confinement, and experiment. In the process we
are also going to find a reliable approach to the light scalar mesons
( $ < 1$ GeV).

The idea is simple and amounts to the observation that valence-quark-pair
creation connects the states of qlQCD to the resonances which in experiment
are seen in elastic and inelastic meson-meson cross sections.
Hence, if we model quark-pair creation in such a way that it theoretically can
be turned on and off, then we are capable to predict the qlQCD states in the
model limit of no-quark-pair creation.
Such a philosophy already underlied an elaborate coupled-channel quark
model \cite{BRMDRR86,BRRD83,Numerical82,BDR80}, designed
to simultaneously describe mesonic bound-state spectra, resonances, and
meson-meson scattering. However, in spite of the model's success to reproduce
a host of experimental data with a very limited number of parameters, it is
clearly not suited as a tool for data analysis, owing to the specific model
choice of the confining $q\bar{q}$ potential, and furthermore the rather
complicated matrix expressions needed to obtain $S$-matrix-related observables.
Therefore, in this work our strategy will be the following.
By replacing the specific confinement part of the Hamiltonian of
Refs.~\cite{BRMDRR86,BRRD83,Numerical82,BDR80} by a more general one, we allow
here for any arbitrary discrete spectrum of ``bare'' $q\bar{q}$ states.
At the same time, by using a simpler form for the coupling potential,
describing the transitions between the $q\bar{q}$ and meson-meson sectors
through $^3P_0$ quark-pair creation, the complexity of the model's scattering
solutions is substantially reduced.
The resulting exact, closed-form formula for the $S$-matrix is subsequently
fitted to the experimental data, i.e., partial-wave cross sections or phase
shifts for meson-meson scattering, by adjusting the model parameters.
Afterwards, we may study the theoretical limit of vanishing
valence-quark-pair creation, described by a sole parameter, which decouples
the bare states from the meson-meson continuum.
The latter are supposed to be equivalent to the states found in qlQCD.
Consequently, the masses of the thus resulting spectrum
could then be compared with those from qlQCD calculations.

The organisation of the present paper is as follows.
In section \ref{BWamplitudes} we develop a model-independent partial-wave
$S$-matrix for the description of elastic and inelastic two-meson scattering.
A so-called ``Resonance-Spectrum Expansion'' (RSE) is discussed in section
(\ref{RSE}).
In section \ref{1threshold} the RSE is compared to the data for elastic
$K\pi$ $S$-wave phase shifts and $P$-wave cross sections in the one-threshold
limit. The complex singularities of the corresponding $S$-matrices
are shown to allow for an easy relation to the $q\bar{q}$ confinement
spectrum. The conclusions are presented in section \ref{finale}.
\clearpage

\section{Breit-Wigner-like scattering amplitudes}
\label{BWamplitudes}

A suitable model in which the communication between the confinement sector of
strong interactions with the two-meson continuum is enabled through
valence-quark-pair creation has been proposed in a series of articles
\cite{BRMDRR86,BRRD83,Numerical82,BDR80}. It allows for the determination
of partial-wave two-meson elastic and inelastic scattering cross sections,
as well as for the search of singularities in the partial-wave
coupled-channel two-meson scattering matrix for all possible valence
flavours. Hence, it enables the calculation of resonances above and bound
states below the lowest threshold, i.e., the meson spectrum.
In the limit of no valence-quark-pair creation one obtains the confinement
spectrum, or bare states, which may be compared to the states of qlQCD.
The model has four parameters and four constituent quark masses.
No distinction is made between up and down quarks. One of the model
parameters parametrises the employed confinement mechanism (harmonic
oscillator), whereas the other three parametrise the communication between the
two distinct sectors of the model, that is, the permanently closed
confinement channels and the meson-meson continuum channels.
The model yields, with one set of four parameters and one set of four
constituent quark masses, good results for heavy quarkonia \cite{BDR80} and
light-meson spectra \cite{BRRD83}, including the scalars \cite{BRMDRR86}, as
well as for two-meson scattering data \cite{BRMDRR86,BRRD83}.

Let us first study the generic form of the scattering matrix for permanently
closed channels coupled to several meson-meson scattering channels,
in a simplified version of the above model.

\subsection{Scattering matrix for several coupled channels}
\label{mccsm}

When we describe quarkonia by wave functions $\psi_{c}$ and two-meson
systems by wave functions $\psi_{f}$, then we obtain for their time evolution
the wave equation

\begin{equation}
\left( E-H_{c}\right)\;\psi_{c}\left(\vec{r}\;\right)\; =\;
V_{t}\;\psi_{f}\left(\vec{r}\;\right)
\;\;\;\xrm{and}\;\;\;
\left( E-H_{f}\right)\;\psi_{f}\left(\vec{r}\;\right)\; =\;
\left[ V_{t}\right]^{T}\;\psi_{c}\left(\vec{r}\;\right)
\;\;\; .
\label{cpldeqna}
\end{equation}

\noindent
Here, $H_{c}$ describes the dynamics of confinement in the interaction region,
$H_{f}$ the dynamics of the scattered particles, and $V_{t}$ the communication
between the two different sectors.

For the dynamics of confinement we understand here that, as a function of
the interquark distance $r$, the resulting quark-antiquark binding forces
grow rapidly outside the interaction region.
Consequently, we must eliminate $\psi_{c}$ from Eqs.~(\ref{cpldeqna}), since it
is vanishing at large distances and thus
{\it unobservable}.
Formally, it is easy to do so. We then obtain the relation

\begin{equation}
\left( E-H_{f}\right)\;\psi_{f}\left(\vec{r}\;\right)\; =\;
\left[ V_{t}\right]^{T}\;
\left( E-H_{c}\right)^{-1}\; V_{t}\;\psi_{f}\left(\vec{r}\;\right)
\;\;\; .
\label{scatteqna}
\end{equation}

By comparison of Eq.~(\ref{scatteqna}) with the usual expressions for
the scattering wave equations, we must conclude that the generalised
potential $V$, which results from the set of coupled equations
(\ref{cpldeqna}), is here given by

\begin{equation}
V\; =\;\left[ V_{t}\right]^{T}\;\left( E-H_{c}\right)^{-1}\; V_{t}
\;\;\; .
\label{generalpot}
\end{equation}

In the momentum representation Eq.~(\ref{generalpot}) takes the form

\begin{equation}
\bra{\vec{p}\,}\; V\;\ket{{\vec{p}\,}'}\; =\;
\bra{\vec{p}\,}\;
\left[ V_{t}\right]^{T}\;\left( E-H_{c}\right)^{-1}\; V_{t}\;
\ket{{\vec{p}\,}'}
\;\;\; .
\label{MSgeneralpot}
\end{equation}

Let us denote the configuration-space representation of the properly
normalised eigensolutions of the operator $H_{c}$ of Eqs.~(\ref{cpldeqna}),
corresponding to the energy eigenvalue $E_{n\ell_{c}}$, by

\begin{equation}
\bracket{\vec{r}\,}{n\ell_{c} m}\; =\;
Y^{(\ell_{c})}_{m}\left(\hat{r}\right)\;
{\cal F}_{n\ell_{c}}\left( r\right)\;\;\; ,
\;\;\xrm{where}\;\;\left\{\begin{array}{l}
\;\xrm{$n=0$, $1$, $2$, $\dots$}\; ,\\ [.3cm]
\;\xrm{$\ell_{c} =0$, $1$, $2$, $\dots$}\; ,\\ [.3cm]
\;\xrm{$m=-\ell_{c}$, $\dots$, $+\ell_{c}$}\; .
\end{array}\right.
\label{complete}
\end{equation}

\noindent
Here, $n$ and $\ell_{c}$ represent the orbital radial and angular
excitations of the quark-antiquark system, respectively.
Hence, when we let the self-adjoint operator $H_{c}$ act to the left,
we obtain for Eq.~(\ref{MSgeneralpot}) the expression

\begin{eqnarray}
\bra{\vec{p}\,}\; V\;\ket{{\vec{p}\,}'} & = &
\sum_{n\ell_{c}  m}\;\bra{\vec{p}\,}\;
\left[ V_{t}\right]^{T}\;
\ket{n\ell_{c}  m}\;\bra{n\ell_{c}  m}\;\left( E-H_{c}\right)^{-1}\; V_{t}\;
\ket{{\vec{p}\,}'}
\nonumber\\ [.3cm] & = &
\sum_{n\ell_{c}  m}\;\bra{\vec{p}\,}\;\left[ V_{t}\right]^{T}\;
\fnd{\ket{n\ell_{c}  m}\;\bra{n\ell_{c}  m}}{E-E_{n\ell_{c} }}
\; V_{t}\;\ket{{\vec{p}\,}'}
\;\;\; .
\label{MSgeneralpot1}
\end{eqnarray}

Evaluation of this equation leads to the Born term of the transition amplitude.

However, now we find it opportune to select the operators $H_{f}$ and $V_{t}$
such that it becomes possible to determine all higher-order terms of the
transition
amplitude. By doing so, we leave no doubt about the analyticity and unitarity
properties of the resulting scattering matrix.
In configuration space we define these operators by

\begin{equation}
H_{f}\; =\; -\frac{1}{2}\;\mu^{-1}\;\nabla^{2}_{r}\;
+\; M_{1}\; +\; M_{2}
\;\;\; ,\;\;\;\xrm{and}\;\;\;
V_{t}\; =\;\fnd{\lambda}{a^{3/2}}\; \bar{V}_{t}\;\delta\left( r-a\right)
\;\;\; ,
\label{Opmodel}
\end{equation}

\noindent
where $\mu$ represents the reduced-mass matrix of the meson-meson system, and
$M_{1,2}$ stand for matrices that contain the masses of the two mesons in each
scattering channel. We limit ourselves to diagonal mass matrices in this work.

The transition potential $V_{t}$ in Eqs.~(\ref{Opmodel}) is, as we shall see
below, a reasonable approximation to the quark-pair-creation transition
potentials described in Ref.~\cite{Bev84}.
It is parametrised by $\lambda$, which determines its intensity, and by
$a$, which stands for the average distance between the interacting
particles (either a quark and an antiquark, or two mesons) where
transitions from one sector to the other take place.
In practice, $a$ should come out of the order of 1 fm,
which is confirmed by adjusting the model parameters to the
experimental data.
$\bar{V}_{t}$ is the matrix that contains the relative intensities for
transitions between the meson-meson and quark-antiquark sector(s).
Notice that we assume here spherical symmetry for all interactions.

With the choices of Eqs.~(\ref{Opmodel}), we obtain for the full (to all orders
in $\lambda$) partial-wave scattering matrix the exact expression

\begin{eqnarray}
\lefteqn{S_{\ell}\left( E\right)\; =\;
\left[ 1\; -\; 2i\fnd{\lambda^{2}}{a}\; p^{-1/2}\; H^{(2)}(a)\;\mu^{1/2}\;
\left[\bar{V}_{t}\right]^{T}
\;\sum_{n=0}^{\infty}\fnd{\abs{{\cal F}_{n\ell_{c}}(a)}^{2}}{E-E_{n\ell_{c}}}
\; \bar{V}_{t}\;\mu^{1/2}\; J(a)\; p^{-1/2}
\right]}
\label{Smat} \\ [.3cm] & & \times
\left[ 1\; +\; 2i\fnd{\lambda^{2}}{a}\; p^{-1/2}\; H^{(1)}(a)\;\mu^{1/2}\;
\left[\bar{V}_{t}\right]^{T}
\;\sum_{n=0}^{\infty}\fnd{\abs{{\cal F}_{n\ell_{c}}(a)}^{2}}{E-E_{n\ell_{c}}}
\; \bar{V}_{t}\;\mu^{1/2}\; J(a)\; p^{-1/2}
\right]^{-1}
\;\;\; ,
\nonumber
\end{eqnarray}

\noindent
where $p$, $\mu$, $J(a)$ and $H^{(1,2)}(a)$ are diagonal matrices throughout
this work, with as many diagonal elements as meson-meson channels considered.
The non-vanishing matrix element

\begin{equation}
p_{i}\; =\;\left[ p\right]_{ii}
\;\;\; ,
\label{linmomi}
\end{equation}

\noindent
represents the relative linear momentum in the centre-of-mass (CM) system of
the $i$-th scattering channel.
The diagonal elements of $J(a)$ and $H^{(1,2)}(a)$ are related to the usual
spherical Bessel and Hankel functions by

\begin{equation}
\left[ J(a)\right]_{ii}\; =\;
p_{i}a\; j_{\ell_{i}}\left( p_{i}a\right)
\;\;\;\xrm{and}\;\;\;
\left[ H^{(1,2)}(a)\right]_{ii}\; =\;
p_{i}a\; h^{(1,2)}_{\ell_{i}}\left( p_{i}a\right)
\;\;\; ,
\label{JH}
\end{equation}

\noindent
where $\ell_{i}$ stands for the relative angular momentum in the $i$-th
scattering channel.

The matrix $\bar{V}_{t}$ contains the coupling constants which are worked out
in Ref.~\cite{BR99b}.
In case only one $q\bar{q}$ channel is considered ($S$-wave meson-meson
scattering for isodoublet and isovector), the matrix $\bar{V}_{t}$ is just
a row vector. Then the expression

\begin{equation}
\sum_{n=0}^{\infty}\fnd{\abs{{\cal F}_{n\ell_{c}}(a)}^{2}}{E-E_{n\ell_{c}}}
\;\;\; 
\label{sum}
\end{equation}

\noindent
is just a real number, that is, a function of the total CM energy $E$.

In case one considers more $q\bar{q}$ channels ($S$-wave meson-meson
scattering for isoscalar coupled $n\bar{n}$ and $s\bar{s}$ channels,
or $P$- and higher-wave scattering), $\bar{V}_{t}$ has as many rows as
$q\bar{q}$ channels.
Then the resonance sum (\ref{sum}) is a real matrix of the size of the
number of $q\bar{q}$ channels.

\subsection{The Resonance-Spectrum Expansion}
\label{RSE}

Expression (\ref{Smat}) contains very little model dependence.
It combines simple kinematics with the experimental observation that
resonances occur in non-exotic scattering processes of mesons.
Not even assumptions are made about possible final-state interactions,
which is expressed by the choice for $H_{f}$ in Eqs.~(\ref{Opmodel}).
Consequently,
since it is not contaminated with model-dependent prejudices,
expression (\ref{Smat}) is extremely
suitable for the analysis of experimental results in two-meson scattering.
Precise determination of the experimental values for $E_{n\ell_{c}}$ and
$\abs{{\cal F}_{n\ell_{c}}(a)}^{2}$ in the resonance sum (\ref{sum}) will give
support to the study of the confinement dynamics and the mechanism of
hadronic decay. Below we shall show how well the procedure works for data
analysis.

In Refs.~\cite{BRMDRR86,BRRD83,BDR80} all possible pseudoscalar-pseudoscalar,
pseudoscalar-vector, and vector-vector scattering channels are coupled,
through $^{3}P_{0}$ nonstrange and strange quark-pair creation, to the
relevant valence quark-antiquark channels. For isovector and isodoublet
flavours one then has one (for $S$-wave scattering) or two (for $P$ and
higher waves) permanently closed channels coupled to many scattering channels.
For the light isoscalars the number of permanently closed channels is doubled,
one channel for $n\bar{n}$ and one for $s\bar{s}$.
The intensities of the relative couplings for the transitions between
permanently closed channels and the various scattering channels are
controlled by flavour independence, which is an observed property of
strong interactions \cite{Abe99}.
This has been worked out in Ref.~\cite{Bev83}, and in some more detail
in Refs.~\cite{BR99b,BR99a}.
As a result, also in Refs.~\cite{BRMDRR86,BRRD83,BDR80} only one overall
parameter is left for all possible transition intensities, which
can be switched on and off. The behaviour of this more complex
model is, especially near the lowest threshold, very similar to the behaviour
of the scattering matrix given in formula (\ref{Smat}).  

In Refs.~\cite{BRMDRR86,BRRD83,BDR80}, a harmonic oscillator with
flavour-independent frequency was chosen for the description of the confinement
dynamics in the permanently closed channels. Hence, by switching off the
overall transition parameter, one obtains the harmonic-oscillator spectrum.
On the other hand, by switching it on, the experimental data for
two-meson scattering are reproduced, as well as bound states like
the Kaon and the $J/\Psi$.
Here, we do not intend to specify the confinement Hamiltonian $H_{c}$ of
Eqs.~(\ref{cpldeqna}), but shall follow a different strategy.
We observe that the expressions ${\cal F}_{n\ell_{c}}\left( a\right)$ in
formula (\ref{Smat}), i.e., the values  of the eigenfunctions of the
confinement operator $H_{c}$ at distance $a$, are $c$-numbers
independent of the total CM energy $E$.
In model \cite{BRMDRR86,BRRD83}, ${\cal F}_{n\ell_{c}}(a)$ and $E_{n\ell_{c}}$
represent the harmonic-oscillator eigenstates and eigenvalues, respectively,
which need only one free parameter, the oscillator frequency.
However, since the confinement mechanism is here supposed to be unknown,
we might as well substitute $\abs{{\cal F}_{n\ell_{c}}\left( a\right)}^{2}$
and $E_{n\ell_{c}}$ by arbitrary non-negative real constants, $B_{n\ell_{c}}$
and $E_{n\ell_{c}}$, to be adjusted to the experimental data, i.e.,

\begin{equation}
\sum_{n=0}^{\infty}\fnd{\abs{{\cal F}_{n\ell_{c}}(a)}^{2}}{E-E_{n\ell_{c}}}
\; =\;\sum_{n=0}^{\infty}\fnd{B_{n\ell_{c}}}{E-E_{n\ell_{c}}}
\;\;\; .
\label{BWsum}
\end{equation}

In practical calculations, one may limit the sum in this ``Resonance-Spectrum
Expansion'' (RSE) to a few ($=N$) resonances only, and
approximate the sum of the remaining terms by a constant,
assuming $E\ll E_{n\ell_{c}}$ for $n>N$.
This way one obtains Breit-Wigner-like expressions.

By redefining $\lambda$ and the $B_{n\ell_{c}}$'s one might take the
above-referred constant equal to $-1$, according to

\begin{equation}
\lambda^{2}\;\sum_{n=0}^{\infty}\fnd{B_{n\ell_{c}}}{E-E_{n\ell_{c}}}
\;\rightarrow\;\lambda^{2}\;\left\{
\sum_{n=0}^{N}\fnd{B_{n\ell_{c}}}{E-E_{n\ell_{c}}}\; -\; 1\right\}
\;\;\; .
\label{BWsumred}
\end{equation}

An alternative approach is to absorb $\lambda^{2}$ into the
$B_{n\ell_{c}}$'s and then separate the relevant terms and the remaining
sum. We shall not follow this strategy, since we want to keep explicit
the dependence on the parameter which provides the communication between the
scattering and confinement sectors.

At this point it is opportune to discuss the model dependence of our
procedure. By the substitution (\ref{BWsum}), any relation to the
quantum numbers of the confinement sector is lost. We just continue
to label the $B_{n\ell_{c}}$'s and $E_{n\ell_{c}}$'s in order to
distinguish them properly. It also implies that a direct reference to the
$^{3}P_{0}$ mechanism is lost.
What is left is just the choice (\ref{Opmodel}) for the remaining
operators, where $H_{f}$ does not even contain final-state interactions, and
$V_{t}$ is a spherically symmetric local approximation to almost
any possible transition mechanism that provides the experimentally observed
OZI-allowed communication between
the confinement and meson-meson scattering sectors.
In summary, the model only assumes that non-exotic meson-meson scattering
is dominated by the coupling to $s$-channel resonances.  
\clearpage

\section{One threshold}
\label{1threshold}

Let us consider the case of one permanently closed channel coupled to one
meson-meson scattering channel.
Using formulae (\ref{Smat}) and (\ref{BWsum}), one deduces
the partial-wave scattering phase shift $\delta_{\ell}(p)$
for elastic meson-meson scattering, which is a function of the
relative momentum in the CM frame,

\begin{equation}
\xrm{cotg}\left(\delta_{\ell}(p)\right)\; =\;
\fnd{n_{\ell}(pa)}{j_{\ell}(pa)}\; -\;
\left[ 2\lambda^{2}\;\mu\; pa\; j^{2}_{\ell}(pa)
\sum_{n=0}^{\infty}\fnd{B_{n\ell_{c}}}{E-E_{n\ell_{c}}}\right]^{-1}
\;\;\; ,
\label{deltaBW}
\end{equation}

\noindent
with $\bar{V}_{t}$ absorbed in $\lambda$.

Notice that the partial-wave phase shifts vanish for
$\lambda\rightarrow 0$, unless one takes at the same time the limit
$E\rightarrow E_{n\ell_{c}}$, which represents therefore the no-interaction
limit in the two-meson system. It is indeed the parameter $\lambda$ that
switches valence-quark-pair creation on and off.

Formula (\ref{deltaBW}) has similar features as standard Breit-Wigner
approximations
\cite{Papa01,Kota00,Work99,Mosh91,Dong98,Fang88,Weinberg82,Orlo84}
for resonant phenomena. However, from the values of $E_{n\ell_{c}}$ and
$B_{n\ell_{c}}$, one cannot read off the positions of the singularities.
At most, one might determine an approximate formula which is good
for small values of $\lambda$, namely

\begin{equation}
E_{n\ell_{c}}\; -\;B_{n\ell_{c}}\;\left[
\sum_{n'\neq n}\fnd{B_{n'\ell_{c}}}{E_{n\ell_{c}}-E_{n'\ell_{c}}}\; -\;
\fnd{i}{2\lambda^{2}\;\mu\; pa\; j_{\ell}(pa)\; h^{(1)}_{\ell}(pa)}
\right]^{-1}
\;\;\; ,
\label{singapprox}
\end{equation}

\noindent
concerning singularities in the complex-energy plane. Their precise locations
can be determined numerically, starting from the approximate values
(\ref{singapprox}).

The values of $E_{n\ell_{c}}$ correspond to the confinement spectrum.
These are the quantities of interest in this work.
Let us consider next the case of $K\pi$ scattering, as an example.

\subsection{\bm{K\pi} \bm{P}-wave scattering}

Since the pion is a rather light particle, we prefer to employ relativistic
kinematics for the relation between the relative linear momentum $p$ in the
$K\pi$ system and the total CM energy $E$, i.e.,

\begin{equation}
E\; =\;\sqrt{p^{2}+m_{\pi}^{2}}\; +\;\sqrt{p^{2}+m_{K}^{2}}
\;\;\;\xrm{or}\;\;\;
p\; =\;\fnd{E}{2}\;\left[
\left\{ 1-\left(\fnd{m_{\pi}+m_{K}}{E}\right)^{2}\right\}
\left\{ 1-\left(\fnd{m_{\pi}-m_{K}}{E}\right)^{2}\right\}\right]^{1/2}
\;\;\; .
\label{pKpi}
\end{equation}

\noindent
Correspondingly, for the reduced mass of the $K\pi$ system we define

\begin{equation}
\mu(E)\; =\frac{1}{2}\fnd{dp^{2}}{dE}\; =\;\fnd{E}{4}\;\left[ 1\; -\;
\left(\fnd{m_{K}^{2}-m_{\pi}^{2}}{E^{2}}\right)^{2}\right]
\;\;\; .
\label{muKpi}
\end{equation}

\begin{figure}[ht]
\normalsize
\begin{center}
\begin{picture}(283.46,293.46)(-50.00,-30.00)
\put(52.81,-5.52){\makebox(0,0)[tc]{0.8}}
\put(118.82,-5.52){\makebox(0,0)[tc]{1.0}}
\put(184.83,-5.52){\makebox(0,0)[tc]{1.2}}
\put(-5.52,49.23){\makebox(0,0)[rc]{100}}
\put(-5.52,98.45){\makebox(0,0)[rc]{200}}
\put(-5.52,147.68){\makebox(0,0)[rc]{300}}
\put(-5.52,196.91){\makebox(0,0)[rc]{400}}
\put(234.34,-5.52){\makebox(0,0)[tr]{$GeV$}}
\put(113.16,-20.07){\makebox(0,0)[tc]{invariant $K\pi$ mass}}
\put(-5.52,250.40){\makebox(0,0)[tr]{GeV$^{-2}$}}
\put(5.52,246.39){\makebox(0,0)[tl]{$\sigma$}}
\put(29.71,8.69){\makebox(0,0){$\bullet$}}
\put(46.21,14.31){\makebox(0,0){$\bullet$}}
\put(59.41,20.48){\makebox(0,0){$\bullet$}}
\put(67.66,48.90){\makebox(0,0){$\bullet$}}
\put(70.96,63.09){\makebox(0,0){$\bullet$}}
\put(74.26,107.28){\makebox(0,0){$\bullet$}}
\put(77.56,145.56){\makebox(0,0){$\bullet$}}
\put(80.86,211.81){\makebox(0,0){$\bullet$}}
\put(84.16,223.61){\makebox(0,0){$\bullet$}}
\put(87.46,194.16){\makebox(0,0){$\bullet$}}
\put(90.77,146.42){\makebox(0,0){$\bullet$}}
\put(94.07,99.44){\makebox(0,0){$\bullet$}}
\put(97.37,76.21){\makebox(0,0){$\bullet$}}
\put(100.67,52.14){\makebox(0,0){$\bullet$}}
\put(103.97,40.73){\makebox(0,0){$\bullet$}}
\put(112.22,23.79){\makebox(0,0){$\bullet$}}
\put(125.42,11.62){\makebox(0,0){$\bullet$}}
\put(138.62,7.67){\makebox(0,0){$\bullet$}}
\put(151.83,4.87){\makebox(0,0){$\bullet$}}
\put(165.03,3.68){\makebox(0,0){$\bullet$}}
\put(178.23,1.24){\makebox(0,0){$\bullet$}}
\put(191.43,1.93){\makebox(0,0){$\bullet$}}
\put(204.63,3.63){\makebox(0,0){$\bullet$}}
\put(217.84,0.26){\makebox(0,0){$\bullet$}}
\end{picture}
\end{center}
\normalsize
\caption[]{Comparison of formula (\ref{deltaBW}) and substitution
(\ref{BWKpiPwave}) with the experimental cross sections for Kaon-pion
$I=\frac{1}{2}$ $P$-wave scattering.
The data are taken from Ref.~\cite{Est78}.}
\label{KpiPwave}
\end{figure}
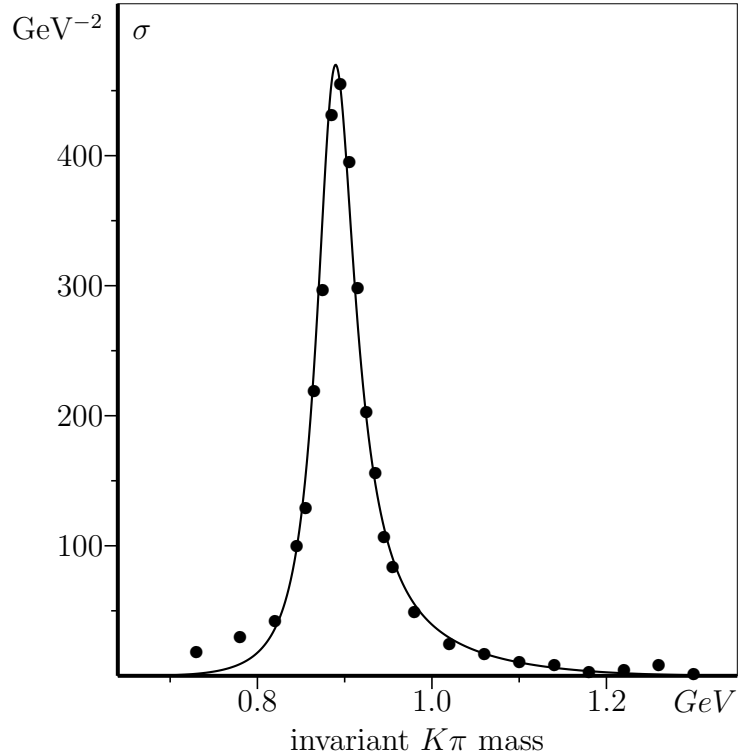

In Fig.~\ref{KpiPwave} we show the result of formula (\ref{deltaBW})
for the cross sections of $I=\frac{1}{2}$ elastic $P$-wave scattering,
for the parameter values $\lambda =0.75$ GeV$^{-3/2}$ and $a=5$ GeV$^{-1}$,
and with the substitution

\begin{equation}
\sum_{n=0}^{\infty}\fnd{B_{n,0}}{E-E_{n,0}}\;
\longrightarrow\;\fnd{0.5}{E-0.945}\; -\; 1
\;\;\;\xrm{GeV}^{\, 2}
\;\;\; ,
\label{BWKpiPwave}
\end{equation}

\noindent
where we neglect possible $\ell_{c}=2$ contributions.


The theoretical curve agrees well with the data.
For the $P$-wave $K\pi$ scattering length, we find here the somewhat too low
result $a_{1}^{1/2}=0.0085\,m^{-1}_{\pi}$,
to be compared to the experimental values taken from Ref.~\cite{Dum83}
(in units of $m^{-1}_{\pi}$), namely
$0.017$ \cite{Lan77}, $0.018\pm 0.002$ \cite{Joh78},
and $0.018$ \cite{Kar80}, or to the chiral-perturbation-theory result
$0.013\pm 0.003$ \cite{Ber91}. Nevertheless, the procedure of substitution
(\ref{BWKpiPwave}) leads to a more than satisfactory description in the
relevant domain of CM energy, thus allowing to read off the value for the bare
$K^{\ast}$(892) mass, i.e., 0.945 GeV.

Moreover, one may inspect the scattering matrix, which follows from
expression (\ref{deltaBW}) after substitution (\ref{BWKpiPwave}), for its
singularities in the complex-energy plane. One finds one pole at

\begin{equation}
0.887\; -\; 0.027\; i
\;\;\;\xrm{GeV}
\;\;\; .
\label{singKpiPwave}
\end{equation}

The relation between the position of the singularity and the Breit-Wigner-like
parameters is lost. However, one gains a simple relation to the confinement
spectrum. Moreover, in the substitution (\ref{BWKpiPwave}) one
may take an arbitrary number of resonances into account.

\begin{figure}[ht]
\Large
\begin{center}
\begin{picture}(425.20,293.46)(-72.50,0.00)
\put(101.15,248.87){\makebox(0,0)[bc]{900}}
\put(202.30,248.87){\makebox(0,0)[bc]{920}}
\put(-8.00,40.14){\makebox(0,0)[rc]{-25}}
\put(-8.00,80.29){\makebox(0,0)[rc]{-20}}
\put(-8.00,120.43){\makebox(0,0)[rc]{-15}}
\put(-8.00,160.58){\makebox(0,0)[rc]{-10}}
\put(-8.00,200.72){\makebox(0,0)[rc]{-5}}
\put(354.02,248.87){\makebox(0,0)[br]{$Re(E)$}}
\put(-8.00,4.01){\makebox(0,0)[br]{$Im(E)$}}
\put(328.73,240.86){\makebox(0,0){$\bullet$}}
\put(324.76,236.10){\makebox(0,0){$\bullet$}}
\put(319.13,230.15){\makebox(0,0){$\bullet$}}
\put(311.17,221.87){\makebox(0,0){$\bullet$}}
\put(300.81,211.29){\makebox(0,0){$\bullet$}}
\put(287.96,198.48){\makebox(0,0){$\bullet$}}
\put(297,198.48){\makebox(0,0)[lc]{\footnotesize $0.30$}}
\put(272.48,183.53){\makebox(0,0){$\bullet$}}
\put(281,183.53){\makebox(0,0)[lc]{\footnotesize $0.35$}}
\put(254.25,166.60){\makebox(0,0){$\bullet$}}
\put(263,166.60){\makebox(0,0)[lc]{\footnotesize $0.40$}}
\put(233.10,147.88){\makebox(0,0){$\bullet$}}
\put(242,147.88){\makebox(0,0)[lc]{\footnotesize $0.45$}}
\put(208.86,127.65){\makebox(0,0){$\bullet$}}
\put(218,127.65){\makebox(0,0)[lc]{\footnotesize $0.50$}}
\put(181.35,106.30){\makebox(0,0){$\bullet$}}
\put(190,106.30){\makebox(0,0)[lc]{\footnotesize $0.55$}}
\put(150.42,84.32){\makebox(0,0){$\bullet$}}
\put(159,84.32){\makebox(0,0)[lc]{\footnotesize $0.60$}}
\put(115.95,62.33){\makebox(0,0){$\bullet$}}
\put(125,62.33){\makebox(0,0)[lc]{\footnotesize $0.65$}}
\put(77.91,41.10){\makebox(0,0){$\bullet$}}
\put(87,41.10){\makebox(0,0)[lc]{\footnotesize $0.70$}}
\put(36.39,21.48){\makebox(0,0){$\bullet$}}
\put(45,21.48){\makebox(0,0)[lc]{\footnotesize $\lambda =0.75$}}
\end{picture}
\end{center}
\normalsize
\normalsize
\caption[]{Complex-energy pole positions of the
scattering matrix, which result from formula (\ref{deltaBW}) and substitution
(\ref{BWKpiPwave}), as a function of the coupling constant $\lambda$.
The point on the real axis corresponds to the bare state ($\lambda =0$).
Units are in MeV.}
\label{Psing}
\end{figure}
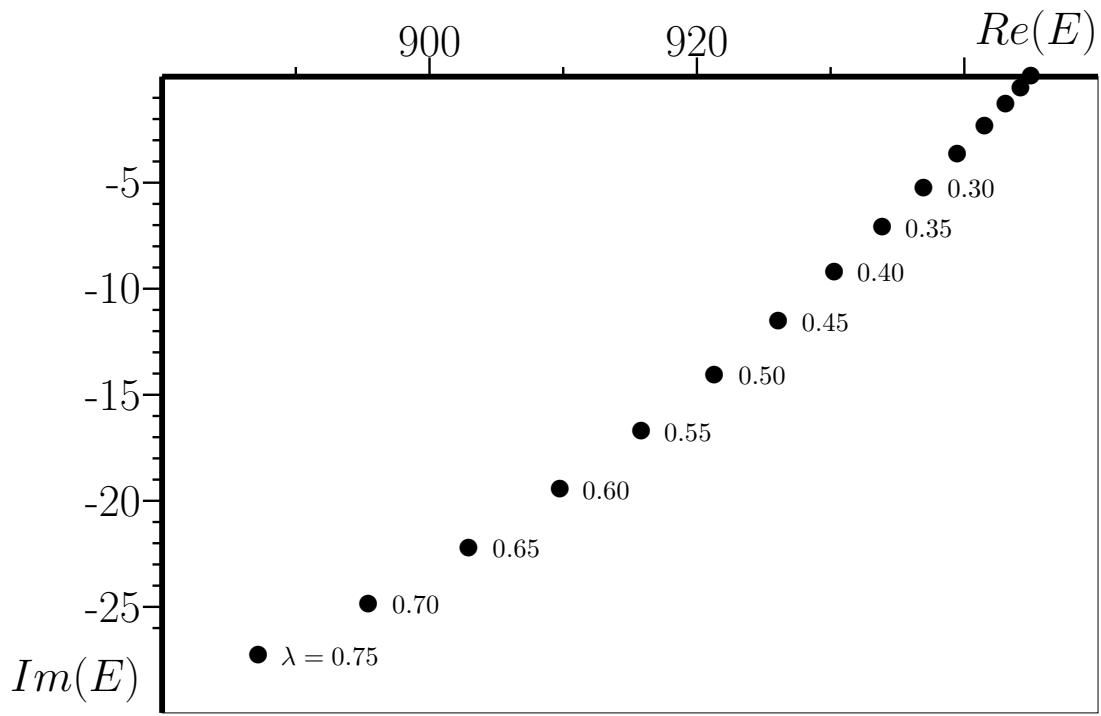

In order to verify that singularity (\ref{singKpiPwave}) stems from the
bare state at 945 MeV, we may stepwise switch off the model
parameter $\lambda$ and inspect the theoretical positions of the
corresponding singularities.
This procedure is shown in Fig.~(\ref{Psing}).


\noindent
It demonstrates beyond any doubt the relation between the singularity
(\ref{singKpiPwave}) and the bare state at 945 MeV.
Notice that the {\it motion} of the singularity for small values of $\lambda$
is perturbative and quadratic in $\lambda$, as indicated by expression
(\ref {singapprox}). However, for larger values of the coupling
constant the singularity positions become more and more nonperturbative.
\clearpage

\subsection{\bm{K\pi} \bm{S}-wave scattering}

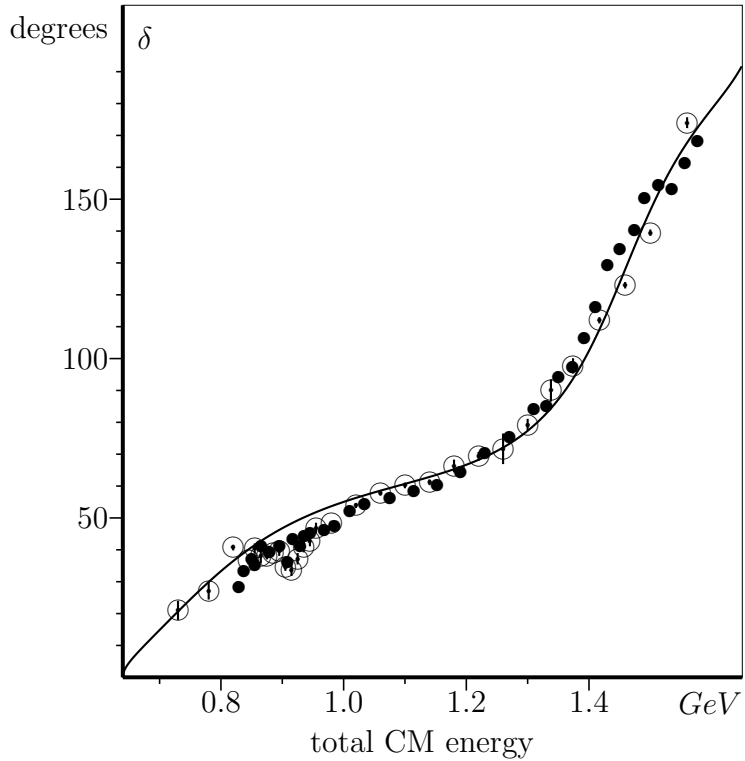
\begin{figure}[ht]
\normalsize
\begin{center}
\begin{picture}(283.46,293.46)(-50.00,-30.00)
\put(37.12,-5.52){\makebox(0,0)[tc]{0.8}}
\put(83.53,-5.52){\makebox(0,0)[tc]{1.0}}
\put(129.93,-5.52){\makebox(0,0)[tc]{1.2}}
\put(176.34,-5.52){\makebox(0,0)[tc]{1.4}}
\put(-5.52,60.33){\makebox(0,0)[rc]{50}}
\put(-5.52,120.67){\makebox(0,0)[rc]{100}}
\put(-5.52,181.00){\makebox(0,0)[rc]{150}}
\put(234.34,-5.52){\makebox(0,0)[tr]{$GeV$}}
\put(113.16,-20.07){\makebox(0,0)[tc]{total CM energy}}
\put(-5.52,250.40){\makebox(0,0)[tr]{degrees}}
\put(5.52,246.39){\makebox(0,0)[tl]{$\delta$}}
\put(20.88,25.34){\makebox(0,0){$\odot$}}
\put(32.48,32.82){\makebox(0,0){$\odot$}}
\put(41.76,49.11){\makebox(0,0){$\odot$}}
\put(47.56,44.40){\makebox(0,0){$\odot$}}
\put(49.88,48.75){\makebox(0,0){$\odot$}}
\put(52.20,45.85){\makebox(0,0){$\odot$}}
\put(54.52,45.97){\makebox(0,0){$\odot$}}
\put(56.84,47.18){\makebox(0,0){$\odot$}}
\put(59.17,47.78){\makebox(0,0){$\odot$}}
\put(61.49,41.87){\makebox(0,0){$\odot$}}
\put(63.81,40.42){\makebox(0,0){$\odot$}}
\put(66.13,44.89){\makebox(0,0){$\odot$}}
\put(68.45,49.35){\makebox(0,0){$\odot$}}
\put(70.77,51.64){\makebox(0,0){$\odot$}}
\put(73.09,56.59){\makebox(0,0){$\odot$}}
\put(78.89,58.40){\makebox(0,0){$\odot$}}
\put(88.17,65.04){\makebox(0,0){$\odot$}}
\put(97.45,69.74){\makebox(0,0){$\odot$}}
\put(106.73,72.64){\makebox(0,0){$\odot$}}
\put(116.01,73.85){\makebox(0,0){$\odot$}}
\put(125.29,80.00){\makebox(0,0){$\odot$}}
\put(134.57,83.62){\makebox(0,0){$\odot$}}
\put(143.85,86.52){\makebox(0,0){$\odot$}}
\put(153.13,95.69){\makebox(0,0){$\odot$}}
\put(161.95,108.60){\makebox(0,0){$\odot$}}
\put(170.30,117.89){\makebox(0,0){$\odot$}}
\put(180.28,135.15){\makebox(0,0){$\odot$}}
\put(190.02,148.42){\makebox(0,0){$\odot$}}
\put(199.54,168.33){\makebox(0,0){$\odot$}}
\put(213.46,209.96){\makebox(0,0){$\odot$}}
\put(43.85,33.79){\makebox(0,0){$\bullet$}}
\put(45.71,39.82){\makebox(0,0){$\bullet$}}
\put(48.72,44.65){\makebox(0,0){$\bullet$}}
\put(49.88,42.23){\makebox(0,0){$\bullet$}}
\put(52.20,49.47){\makebox(0,0){$\bullet$}}
\put(55.22,47.06){\makebox(0,0){$\bullet$}}
\put(59.17,49.47){\makebox(0,0){$\bullet$}}
\put(62.18,43.44){\makebox(0,0){$\bullet$}}
\put(64.27,51.89){\makebox(0,0){$\bullet$}}
\put(67.05,49.47){\makebox(0,0){$\bullet$}}
\put(68.45,53.09){\makebox(0,0){$\bullet$}}
\put(70.77,54.30){\makebox(0,0){$\bullet$}}
\put(76.10,55.51){\makebox(0,0){$\bullet$}}
\put(80.05,56.71){\makebox(0,0){$\bullet$}}
\put(85.85,62.75){\makebox(0,0){$\bullet$}}
\put(91.42,65.16){\makebox(0,0){$\bullet$}}
\put(100.93,67.57){\makebox(0,0){$\bullet$}}
\put(109.98,69.99){\makebox(0,0){$\bullet$}}
\put(118.79,72.40){\makebox(0,0){$\bullet$}}
\put(127.61,77.23){\makebox(0,0){$\bullet$}}
\put(136.89,84.47){\makebox(0,0){$\bullet$}}
\put(146.17,90.50){\makebox(0,0){$\bullet$}}
\put(155.45,101.36){\makebox(0,0){$\bullet$}}
\put(160.09,102.57){\makebox(0,0){$\bullet$}}
\put(164.73,113.43){\makebox(0,0){$\bullet$}}
\put(170.07,117.05){\makebox(0,0){$\bullet$}}
\put(174.48,127.91){\makebox(0,0){$\bullet$}}
\put(178.66,139.97){\makebox(0,0){$\bullet$}}
\put(183.30,155.66){\makebox(0,0){$\bullet$}}
\put(187.94,161.69){\makebox(0,0){$\bullet$}}
\put(193.50,168.93){\makebox(0,0){$\bullet$}}
\put(197.22,181.00){\makebox(0,0){$\bullet$}}
\put(202.55,185.83){\makebox(0,0){$\bullet$}}
\put(207.66,184.62){\makebox(0,0){$\bullet$}}
\put(212.53,194.27){\makebox(0,0){$\bullet$}}
\put(217.40,202.72){\makebox(0,0){$\bullet$}}
\end{picture}
\end{center}
\normalsize
\caption[]{Comparison of formula (\ref{deltaBW}) and substitution
(\ref{BWKpiSwave}) with the experimental phase shifts for Kaon-pion
$I=\frac{1}{2}$ $S$-wave scattering.
The data are taken from Refs.~\cite{Est78,Est79} ($\odot$) and 
\cite{Ast88} ($\bullet$).}
\label{KpiSwave}
\end{figure}

In Fig.~\ref{KpiSwave} we show the result of formula (\ref{deltaBW})
for the phase shifts of $I=\frac{1}{2}$ elastic $S$-wave scattering,
for the values $\lambda =0.75$ GeV$^{-3/2}$ and $a=3.2$ GeV$^{-1}$,
and with the substitution

\begin{equation}
\sum_{n=0}^{\infty}\fnd{B_{n,1}}{E-E_{n,1}}\;
\longrightarrow\;
\fnd{1.0}{E-1.31}\; +\; \fnd{0.2}{E-1.69}\; -\; 1
\;\;\;\xrm{GeV}^{\: 2}
\;\;\; .
\label{BWKpiSwave}
\end{equation}


Also for $S$-wave scattering we find good agreement between the theoretical
curve and the data. The difference between $P$ and $S$ wave for the radius
$a$ of the transition potential agrees well with the corresponding difference
for the full pseudoscalar-pseudoscalar transition
potentials depicted in Figs.~7 and 8 of Ref.~\cite{Bev84}, respectively. 
For the $S$-wave $K\pi$ scattering length we find here
$a_{0}^{1/2}=0.22\,m^{-1}_{\pi}$,
to be compared to the experimental values taken from Ref.~\cite{Dum83}
(in units of $m^{-1}_{\pi}$), i.e.,
$0.33\pm 0.01$ \cite{Est78}, $0.237$ \cite{Lan77},
$0.240\pm 0.002$ \cite{Joh78}, $0.13\pm 0.09$ \cite{Kar80},
and $0.22\pm 0.04$ \cite{Mat74}, or to the chiral-perturbation-theory result
$0.17\pm 0.02$ \cite{Ber91}.

From Eq.~(\ref{BWKpiSwave}) one reads for the lowest $J^{P}=0^{+}$
isodoublet eigenstates of confinement the bare masses

\begin{equation}
E_{0,1}\; =\; 1.31\;\;\xrm{GeV}
\;\;\;\xrm{and}\;\;\;
E_{1,1}\; =\; 1.69\;\;\xrm{GeV}
\;\;\; .
\label{EKpiSwave}
\end{equation}

\noindent
However, when we search formula (\ref{deltaBW}), after substitution
(\ref{BWKpiSwave}), for singularities, then we find, besides the two
corresponding singularities at

\begin{equation}
(1.458\; -\; 0.118\; i)\;\;\xrm{GeV}
\;\;\;\xrm{and}\;\;\;
(1.713\; -\; 0.019\; i)\;\;\xrm{GeV}
\;\;\; ,
\label{singKpiSwave}
\end{equation}

\noindent
also an additional one at

\begin{equation}
(0.714\; -\; 0.228\; i)\;\;\xrm{GeV}
\;\;\; .
\label{kappa}
\end{equation}


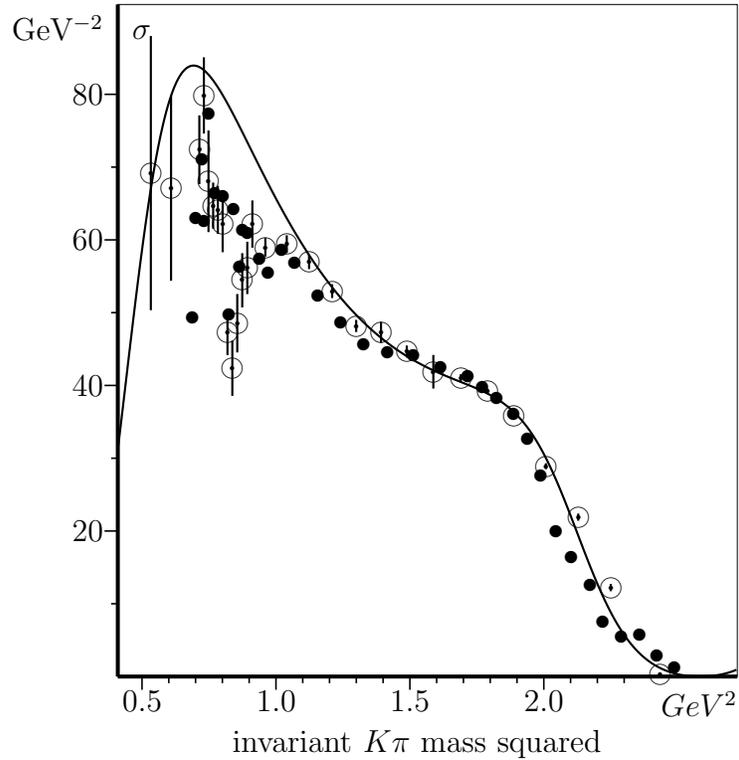
\begin{figure}[ht]
\normalsize
\begin{center}
\begin{picture}(283.46,293.46)(-50.00,-30.00)
\put(9.16,-5.52){\makebox(0,0)[tc]{0.5}}
\put(59.82,-5.52){\makebox(0,0)[tc]{1.0}}
\put(110.48,-5.52){\makebox(0,0)[tc]{1.5}}
\put(161.14,-5.52){\makebox(0,0)[tc]{2.0}}
\put(-5.52,55.10){\makebox(0,0)[rc]{20}}
\put(-5.52,110.20){\makebox(0,0)[rc]{40}}
\put(-5.52,165.30){\makebox(0,0)[rc]{60}}
\put(-5.52,220.40){\makebox(0,0)[rc]{80}}
\put(234.34,-5.52){\makebox(0,0)[tr]{$GeV^{2}$}}
\put(113.16,-20.07){\makebox(0,0)[tc]{invariant $K\pi$ mass squared}}
\put(-5.52,250.40){\makebox(0,0)[tr]{GeV$^{-2}$}}
\put(5.52,246.39){\makebox(0,0)[tl]{$\sigma$}}
\put(12.49,190.59){\makebox(0,0){$\odot$}}
\put(20.14,184.94){\makebox(0,0){$\odot$}}
\put(30.84,199.43){\makebox(0,0){$\odot$}}
\put(32.57,219.98){\makebox(0,0){$\odot$}}
\put(34.31,187.48){\makebox(0,0){$\odot$}}
\put(36.07,178.27){\makebox(0,0){$\odot$}}
\put(37.86,176.66){\makebox(0,0){$\odot$}}
\put(39.66,171.49){\makebox(0,0){$\odot$}}
\put(41.48,130.22){\makebox(0,0){$\odot$}}
\put(43.33,116.72){\makebox(0,0){$\odot$}}
\put(45.19,133.75){\makebox(0,0){$\odot$}}
\put(47.08,150.01){\makebox(0,0){$\odot$}}
\put(48.98,154.69){\makebox(0,0){$\odot$}}
\put(50.91,171.29){\makebox(0,0){$\odot$}}
\put(55.81,162.53){\makebox(0,0){$\odot$}}
\put(63.91,163.72){\makebox(0,0){$\odot$}}
\put(72.34,156.98){\makebox(0,0){$\odot$}}
\put(81.10,145.91){\makebox(0,0){$\odot$}}
\put(90.17,132.72){\makebox(0,0){$\odot$}}
\put(99.58,130.26){\makebox(0,0){$\odot$}}
\put(109.30,123.04){\makebox(0,0){$\odot$}}
\put(119.35,115.39){\makebox(0,0){$\odot$}}
\put(129.73,113.10){\makebox(0,0){$\odot$}}
\put(139.89,108.13){\makebox(0,0){$\odot$}}
\put(149.78,98.78){\makebox(0,0){$\odot$}}
\put(161.94,79.62){\makebox(0,0){$\odot$}}
\put(174.18,60.33){\makebox(0,0){$\odot$}}
\put(186.47,33.68){\makebox(0,0){$\odot$}}
\put(205.07,0.79){\makebox(0,0){$\odot$}}
\put(28.13,135.48){\makebox(0,0){$\bullet$}}
\put(29.48,173.22){\makebox(0,0){$\bullet$}}
\put(31.70,195.33){\makebox(0,0){$\bullet$}}
\put(32.57,172.28){\makebox(0,0){$\bullet$}}
\put(34.31,212.89){\makebox(0,0){$\bullet$}}
\put(36.60,182.48){\makebox(0,0){$\bullet$}}
\put(39.66,181.67){\makebox(0,0){$\bullet$}}
\put(42.03,136.83){\makebox(0,0){$\bullet$}}
\put(43.70,176.56){\makebox(0,0){$\bullet$}}
\put(45.94,154.68){\makebox(0,0){$\bullet$}}
\put(47.08,168.86){\makebox(0,0){$\bullet$}}
\put(48.98,167.54){\makebox(0,0){$\bullet$}}
\put(53.44,157.67){\makebox(0,0){$\bullet$}}
\put(56.80,152.49){\makebox(0,0){$\bullet$}}
\put(61.86,161.35){\makebox(0,0){$\bullet$}}
\put(66.83,156.38){\makebox(0,0){$\bullet$}}
\put(75.59,143.70){\makebox(0,0){$\bullet$}}
\put(84.24,133.76){\makebox(0,0){$\bullet$}}
\put(92.96,125.45){\makebox(0,0){$\bullet$}}
\put(101.98,122.34){\makebox(0,0){$\bullet$}}
\put(111.78,121.22){\makebox(0,0){$\bullet$}}
\put(121.92,116.76){\makebox(0,0){$\bullet$}}
\put(132.37,113.40){\makebox(0,0){$\bullet$}}
\put(137.72,109.10){\makebox(0,0){$\bullet$}}
\put(143.15,105.00){\makebox(0,0){$\bullet$}}
\put(149.50,99.28){\makebox(0,0){$\bullet$}}
\put(154.82,89.74){\makebox(0,0){$\bullet$}}
\put(159.93,75.81){\makebox(0,0){$\bullet$}}
\put(165.69,54.61){\makebox(0,0){$\bullet$}}
\put(171.52,45.11){\makebox(0,0){$\bullet$}}
\put(178.63,34.51){\makebox(0,0){$\bullet$}}
\put(183.44,20.31){\makebox(0,0){$\bullet$}}
\put(190.44,15.01){\makebox(0,0){$\bullet$}}
\put(197.23,15.52){\makebox(0,0){$\bullet$}}
\put(203.81,7.71){\makebox(0,0){$\bullet$}}
\put(210.47,3.04){\makebox(0,0){$\bullet$}}
\end{picture}
\end{center}
\normalsize
\caption[]{Kaon-pion $I=\frac{1}{2}$ $S$-wave cross sections
deduced from the phase shifts of Fig.~\ref{KpiSwave}.}
\label{crKpiS}
\end{figure}

In Fig.~\ref{crKpiS} we depict the transformation of the theoretical and
experimental phase shifts of Fig.~\ref{KpiSwave} into the $I=\frac{1}{2}$
elastic partial $S$-wave cross sections, which show a clear signal at about
830 {MeV}, with a width of some 500 {MeV}, as possibly also seen in a very
recent experiment by the E791 collaboration \cite{Goebel00}.
Notice, however, that neither the theoretical, nor the experimental cross
sections exhibit a dip, characteristic for Breit-Wigner distributions.

The singularity (\ref{kappa}), which in model \cite{BRMDRR86} comes out at
$(0.727\; -\; 0.263\; i)\;\;\xrm{GeV}$, is
interpreted as the isodoublet partner $K_{0}^{\ast}(727)$
of the $f_{0}(400-1200)$. 
A corresponding resonance has also been reported in several other works
\cite{Bla100,Bla200,Sca00,Jamin00,Ishida00,Bab00,Mar99,Nap99,Ishida99,Far99,
Bla98,Oller99,Oller98,Ishida97} in more recent years.
The preliminary experimental result of the E791 collaboration reported
in Ref.~\cite{Goebel00} awaits further confirmation.

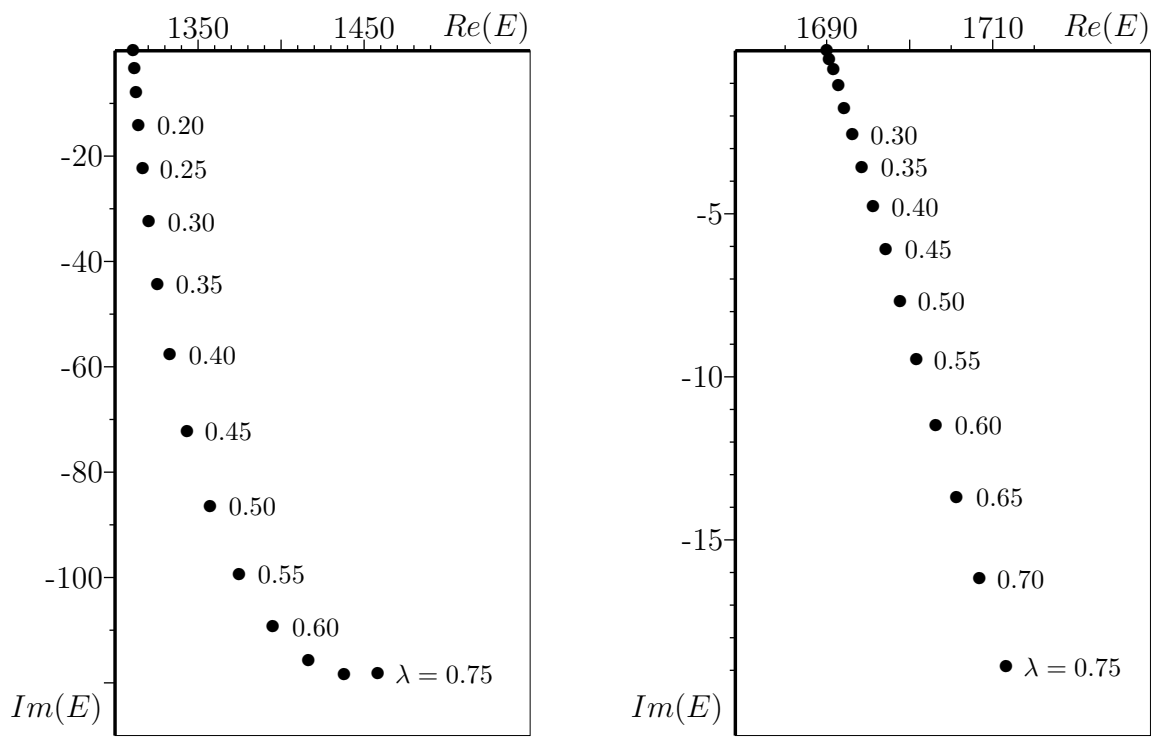
\begin{figure}[ht]
\normalsize
\begin{center}
\begin{picture}(198.43,293.46)(-42.00,0.00)
\put(31.40,263.87){\makebox(0,0)[bc]{1350}}
\put(94.21,263.87){\makebox(0,0)[bc]{1450}}
\put(-4.64,59.82){\makebox(0,0)[rc]{-100}}
\put(-4.64,99.71){\makebox(0,0)[rc]{-80}}
\put(-4.64,139.59){\makebox(0,0)[rc]{-60}}
\put(-4.64,179.47){\makebox(0,0)[rc]{-40}}
\put(-4.64,219.35){\makebox(0,0)[rc]{-20}}
\put(157.01,262.0){\makebox(0,0)[br]{$Re(E)$}}
\put(-4.64,4.01){\makebox(0,0)[br]{$Im(E)$}}
\put(6.91,259.23){\makebox(0,0){$\bullet$}}
\put(7.35,252.25){\makebox(0,0){$\bullet$}}
\put(7.98,243.28){\makebox(0,0){$\bullet$}}
\put(8.98,230.72){\makebox(0,0){$\bullet$}}
\put(10.49,214.37){\makebox(0,0){$\bullet$}}
\put(12.75,194.42){\makebox(0,0){$\bullet$}}
\put(16.02,170.69){\makebox(0,0){$\bullet$}}
\put(20.73,143.97){\makebox(0,0){$\bullet$}}
\put(27.26,115.06){\makebox(0,0){$\bullet$}}
\put(36.05,86.54){\makebox(0,0){$\bullet$}}
\put(46.98,60.82){\makebox(0,0){$\bullet$}}
\put(59.66,40.88){\makebox(0,0){$\bullet$}}
\put(73.17,28.32){\makebox(0,0){$\bullet$}}
\put(86.61,22.73){\makebox(0,0){$\bullet$}}
\put(99.36,23.33){\makebox(0,0){$\bullet$}}
\put(16,230.72){\makebox(0,0)[lc]{\footnotesize $0.20$}}
\put(17,214.37){\makebox(0,0)[lc]{\footnotesize $0.25$}}
\put(20,194.4){\makebox(0,0)[lc]{\footnotesize $0.30$}}
\put(23,170.69){\makebox(0,0)[lc]{\footnotesize $0.35$}}
\put(28,143.97){\makebox(0,0)[lc]{\footnotesize $0.40$}}
\put(34,115.06){\makebox(0,0)[lc]{\footnotesize $0.45$}}
\put(43,86.54){\makebox(0,0)[lc]{\footnotesize $0.50$}}
\put(54,60.82){\makebox(0,0)[lc]{\footnotesize $0.55$}}
\put(67,40.88){\makebox(0,0)[lc]{\footnotesize $0.60$}}
\put(106,23.33){\makebox(0,0)[lc]{\footnotesize $\lambda =0.75$}}
\end{picture}
\hspace{1cm}
\begin{picture}(198.43,293.46)(-42.00,0.00)
\put(34.53,263.87){\makebox(0,0)[bc]{1690}}
\put(97.34,263.87){\makebox(0,0)[bc]{1710}}
\put(-4.64,74.07){\makebox(0,0)[rc]{-15}}
\put(-4.64,135.79){\makebox(0,0)[rc]{-10}}
\put(-4.64,197.51){\makebox(0,0)[rc]{-5}}
\put(157.01,262){\makebox(0,0)[br]{$Re(E)$}}
\put(-4.64,4.01){\makebox(0,0)[br]{$Im(E)$}}
\put(34.54,259.23){\makebox(0,0){$\bullet$}}
\put(35.48,255.53){\makebox(0,0){$\bullet$}}
\put(37.05,251.83){\makebox(0,0){$\bullet$}}
\put(38.94,245.65){\makebox(0,0){$\bullet$}}
\put(41.14,237.01){\makebox(0,0){$\bullet$}}
\put(44.28,227.14){\makebox(0,0){$\bullet$}}
\put(47.73,214.79){\makebox(0,0){$\bullet$}}
\put(52.13,199.98){\makebox(0,0){$\bullet$}}
\put(56.84,183.93){\makebox(0,0){$\bullet$}}
\put(62.18,164.18){\makebox(0,0){$\bullet$}}
\put(68.46,141.96){\makebox(0,0){$\bullet$}}
\put(75.68,117.27){\makebox(0,0){$\bullet$}}
\put(83.53,90.11){\makebox(0,0){$\bullet$}}
\put(92.32,59.25){\makebox(0,0){$\bullet$}}
\put(102.37,25.92){\makebox(0,0){$\bullet$}}
\put(51,227.14){\makebox(0,0)[lc]{\footnotesize $0.30$}}
\put(55,214.79){\makebox(0,0)[lc]{\footnotesize $0.35$}}
\put(59,199.98){\makebox(0,0)[lc]{\footnotesize $0.40$}}
\put(64,183.93){\makebox(0,0)[lc]{\footnotesize $0.45$}}
\put(69,164.18){\makebox(0,0)[lc]{\footnotesize $0.50$}}
\put(75,141.96){\makebox(0,0)[lc]{\footnotesize $0.55$}}
\put(83,117.27){\makebox(0,0)[lc]{\footnotesize $0.60$}}
\put(91,90.11){\makebox(0,0)[lc]{\footnotesize $0.65$}}
\put(99,59.25){\makebox(0,0)[lc]{\footnotesize $0.70$}}
\put(109,25.92){\makebox(0,0)[lc]{\footnotesize $\lambda =0.75$}}
\end{picture}
\end{center}
\normalsize
\caption[]{Complex-energy pole positions of the
scattering matrix, which result from formula (\ref{deltaBW}) and substitution
(\ref{BWKpiSwave}), as a function of the coupling constant $\lambda$.
The points on the real axes correspond to the bare states ($\lambda =0$).
Units are in MeV.}
\label{Ssing1}
\end{figure}

The singularities (\ref{singKpiSwave}) correspond to the
ground state ($n=0$, $\ell_{c}=1$) at $E_{0,1}$ and the first radially
excited state ($n=1$, $\ell_{c}=1$) at $E_{1,1}$ of the confinement
spectrum, respectively.
By reducing the value of the coupling constant $\lambda$, the singularities
(\ref{singKpiSwave}) move towards $E_{0,1}$ and $E_{1,1}$, respectively, which
can be most clearly understood from expression (\ref{singapprox}).
The complex-energy singularities of the partial $S$-wave scattering matrix
(\ref{Smat}), resulting from stepwise reducing the theoretical coupling
constant $\lambda$ in formula (\ref{deltaBW}), are depicted in
Fig.~(\ref{Ssing1}).

\noindent
It clearly demonstrates the relation between the singularities
(\ref{singKpiSwave}) and the bare states (\ref{EKpiSwave}).
Nonperturbative effects for larger values of the coupling constant
can in particular be observed for the lower of the two resonances.
First- (or second-) order perturbative calculations would result in
completely different positions for the singularities corresponding
to the model value of the coupling constant. Especially the real part
of the {\it mass shift} \/is strongly affected by higher-order corrections.
In a recent $K$-matrix analysis \cite{Ani97}, as well as in
a covariant quarkonium model \cite{Koll00}, it is indeed also found that
the bare state might be appreciably lower than the central resonance
position for the $K^{\ast}_{0}(1430)$.

\begin{figure}[ht]
\normalsize
\begin{center}
\begin{picture}(141.73,293.46)(-30.50,0.00)
\put(20.30,269.53){\makebox(0,0)[bc]{0.66}}
\put(60.90,269.53){\makebox(0,0)[bc]{0.70}}
\put(-3.37,66.54){\makebox(0,0)[rc]{-1.5}}
\put(-3.37,133.08){\makebox(0,0)[rc]{-1.0}}
\put(-3.37,199.62){\makebox(0,0)[rc]{-0.5}}
\put(111.65,269.53){\makebox(0,0)[br]{$Re(E)$}}
\put(-3.37,4.01){\makebox(0,0)[br]{$Im(E)$}}
\put(75.01,235.83){\makebox(0,0){$\bullet$}}
\put(68,235.83){\makebox(0,0)[rc]{\footnotesize $\lambda =0.75$}}
\put(83.84,230.25){\makebox(0,0){$\bullet$}}
\put(91.35,223.60){\makebox(0,0){$\bullet$}}
\put(96.53,215.62){\makebox(0,0){$\bullet$}}
\put(98.35,206.11){\makebox(0,0){$\bullet$}}
\put(96.53,194.95){\makebox(0,0){$\bullet$}}
\put(91.65,182.15){\makebox(0,0){$\bullet$}}
\put(85.16,167.80){\makebox(0,0){$\bullet$}}
\put(92,167.80){\makebox(0,0)[lc]{\footnotesize $0.40$}}
\put(78.05,152.00){\makebox(0,0){$\bullet$}}
\put(85,152.00){\makebox(0,0)[lc]{\footnotesize $0.35$}}
\put(70.14,134.56){\makebox(0,0){$\bullet$}}
\put(77,134.56){\makebox(0,0)[lc]{\footnotesize $0.30$}}
\put(61.00,114.90){\makebox(0,0){$\bullet$}}
\put(68,114.90){\makebox(0,0)[lc]{\footnotesize $0.25$}}
\put(50.04,91.93){\makebox(0,0){$\bullet$}}
\put(57,91.93){\makebox(0,0)[lc]{\footnotesize $0.20$}}
\put(36.54,63.55){\makebox(0,0){$\bullet$}}
\put(43,63.55){\makebox(0,0)[lc]{\footnotesize $0.15$}}
\put(19.18,25.06){\makebox(0,0){$\bullet$}}
\put(26,25.06){\makebox(0,0)[lc]{\footnotesize $\lambda =0.10$}}
\end{picture}
\end{center}
\normalsize
\caption[]{Complex-energy pole positions of the
scattering matrix, which result from formula (\ref{deltaBW}) and substitution
(\ref{BWKpiSwave}), as a function of the coupling constant $\lambda$.
Units are in GeV.}
\label{Ssing0}
\end{figure}
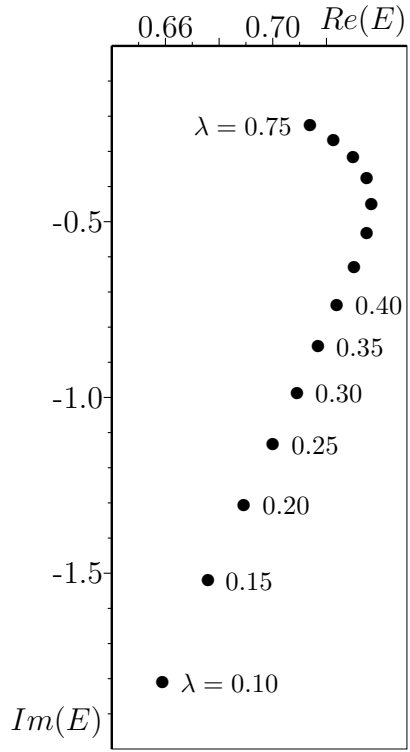

The $K_{0}^{\ast}(727)$ singularity (\ref{kappa}) does not have a direct
relation to any of the states stemming from the confinement mechanism.
This fact is also most clearly demonstrated by the theoretical process of
decoupling the $\pi K$ sector from the strange-nonstrange quarkonium
sector, as depicted in Fig.~(\ref{Ssing0}).

The singularity (\ref{kappa}) acquires a larger imaginary part
when $\lambda$ is reduced, thus describing a state with ever increasing width,
which is a highly nonperturbative phenomenon.
For vanishing coupling constant, one observes that the $K_{0}^{\ast}(727)$
disappears into the scattering background.
This is a very important observation, since it implies that the
$K_{0}^{\ast}(727)$ singularity is a consequence of the transition
mechanism that provides the communication between the $\pi K$ sector
and the strange-nonstrange quarkonium sector.
No such phenomena are observed for $P$- and higher-wave meson-meson scattering.
So we may conclude that a corresponding effect is screened 
from observation by the centrifugal barrier in Eq.~(\ref{scatteqna}).
Consequently, the mechanism of valence-quark-pair creation appears to be
more important in $S$-wave meson-meson scattering.

The absence of a direct relation between the $K_{0}^{\ast}(727)$
singularity and the nonstrange-strange quarkonium sector has for the
first time been demonstrated in Ref.~\cite{BRMDRR86}, where similar
singularities are shown to describe the properties of the $f_{0}(400-1200)$,
$f_{0}(980)$, and $a_{0}(980)$ resonances, thereby resolving two important
issues: the nature of the light scalar mesons and the completion of
the light scalar nonet.

The positions of the various singularities for $\pi K$ $S$-wave scattering
in the complex-energy plane as a function of $\lambda$ can of course 
be obtained from the analyticity properties of the scattering matrix $S(E)$
(\ref{Smat}), or, as no approximations are made in the determination
of $S(E)$, directly from the dynamical equations (\ref{cpldeqna}) and
(\ref{Opmodel}).
In an extensive study on multichannel scattering with permanently closed
channels \cite{DHM76}, resonances like the here described $K^{\ast}_{0}(727)$
are distinguished from the {\it fundamental} \/resonances related to
the bound states of $H_{c}$, and referred to as
{\it hadron molecular states}. The idea of {\it meson molecules} \/has been
worked out in several papers \cite{JW91,JWIs91}, as a possible explanation
for the light scalar mesons $a_0$(980) and $f_0$(980).
However, the term {\it molecule} \/gives the wrong impression that the
$q\bar{q}$ component has no importance for these states.
From the dynamical equations (\ref{cpldeqna}) one may determine
the contribution of the strange-nonstrange quarkonium sector to the
wave functions for any of the states under the $K^{\ast}_{0}(727)$
resonance.

The theoretical distribution of Fig.~\ref{crKpiS} could well be used
for experimental analysis, by optimising the adjustment of the here
proposed parameters to the data.
Especially the absence of the dip in the cross section is well accounted
for in formula (\ref{deltaBW}) with substitution (\ref{BWKpiSwave}).
It has moreover the advantage that a direct relation exists between the
theoretical distribution and the phase shifts and scattering matrices,
which enables the precise location of the singularity associated with the
$K_{0}^{\ast}(727)$.
\clearpage

\section{Conclusions}
\label{finale}

We have shown that the RSE parameters $E_{n\ell_{c}}$ of formula
(\ref{BWsum}) relate experiment to qlQCD calculations of
hadron masses better than the usual central Breit-Wigner positions
of resonances do.
This is not only a consequence of the potentially large mass shifts due to
hadronic decay, but also owing to the nice feature of the RSE procedure
that $S$-matrix singularities not originating from genuine confinement
can easily be isolated.
As an important application of the RSE, it is shown here, through the example
of the $K_{0}^{\ast}(727)$, how the light scalar mesons can be described by
$S$-matrix singularities which are not directly related to the ground states
of the $J^{P}=0^{+}$ confinement spectrum.

The RSE parameters $B_{n\ell_{c}}$ of formula (\ref{BWsum}) incorporate
the unknowns of hadronic decay processes.
Empirical knowledge of these parameters will certainly give a substantial
contribution to the detailed study of hadronic decay at low and
intermediate energies.

With respect to the quantitative conclusions of the present work, a word of
caution is in place.
Since expression (\ref{Smat}) allows for it, we may also inspect the effects
of higher thresholds. This has been carried out for harmonic-oscillator
confinement in the model of
Refs.~\cite{BRMDRR86,BRRD83,Numerical82,BDR80,Bev84,BR99b},
with the following results: for $P$-wave non-exotic two-meson scattering,
both with $J^{PC}=0^{-+}$ and $J^{PC}=1^{--}$, the real parts of the
singularities which correspond to the ground states at $E_{0,0}$ of the
confinement spectrum come out some 300--400 MeV below $E_{0,0}$.
For the higher radial excitations these shifts are considerably smaller.
For $S$-waves the shifts are also smaller for the ground states and,
moreover, in the positive direction.
When we compare these findings with the one-threshold results shown above,
we must conclude that higher thresholds should be taken into account for a
more quantitative determination of the confinement and decay mechanisms.
In particular, the second singularity of formula (\ref{singKpiSwave}) might
come out rather displaced, if higher thresholds are accounted for.
This will be investigated in future work.
\vspace{0.3cm}

{\it Acknowledgements.} We are indebted to Jeffrey Appel for drawing our
attention to the need for a more phenomenological approach to data analysis
in coupled-channel models for confinement and hadronic decay.
We also wish to thank Alexander A.\ Osipov, Brigitte Hiller, Alex Blin
and Hans Walliser for useful discussions and suggestions.
This work is partly supported by the
{\it Funda\c{c}\~{a}o para a Ci\^{e}ncia e a Tecnologia}
of the {\it Minist\'{e}rio da
Ci\^{e}ncia e da Tecnologia} \/of Portugal,
under contract numbers
PESO/\-P/\-PRO/\-15127/\-99,
POCTI/\-35304/\-FIS/\-2000 and CERN/\-P/\-FIS/\-40119/\-2000.


\clearpage

\begin{thebibliography}{45}
\bibitem{Weingarten94}
F.~Butler, H.~Chen, J.~Sexton, A.~Vaccarino and D.~Weingarten,
Nucl.\ Phys.\ {\bf B430}, 179 (1994)
[hep-lat/9405003].

\bibitem{Weingarten83}
D.~Weingarten,
Nucl.\ Phys.\ {\bf B215}, 1 (1983).

\bibitem{Hamber82} 
H.~Hamber, E.~Marinari, G.~Parisi and C.~Rebbi,
Phys.\ Lett.\ {\bf B108}, 314 (1982).

\bibitem{Hamber81} 
H.~Hamber and G.~Parisi,
Phys.\ Rev.\ Lett.\  {\bf 47}, 1792 (1981).

\bibitem{BDR83}
E.~van Beveren, C.~Dullemond and T.~A.~Rijken,
Z.\ Phys.\ {\bf C19}, 275 (1983).

\bibitem{Ali0105015}  
A.~A.~Khan {\it et al.}  [CP-PACS Collaboration],
hep-lat/0105015.

\bibitem{Aoki99} 
S.~Aoki {\it et al.}  [CP-PACS Collaboration],
Phys.\ Rev.\ Lett.\  {\bf 84}, 238 (2000)
[hep-lat/9904012].

\bibitem{Kuramashi99} 
Yoshinobu~Kuramashi for the CP-PACS Collaboration,
Proceedings of the Meeting of Particles and Fields of the American
Physical Society (DPF99), Los Angeles, CA, 5-9 Jan 1999
[hep-lat/9904003].

\bibitem{LW99}
W.~Lee and D.~Weingarten,
Phys.\ Rev.\ {\bf D61}, 014015 (2000)
[hep-lat/9910008];

Ibid.,
hep-lat/9805029;

Ibid.,
Nucl.\ Phys.\ Proc.\ Suppl.\  {\bf 53}, 236 (1997)
[hep-lat/9608071];

D.~Weingarten, private communication.

\bibitem{BRMDRR86}                 
E.~van~Beveren, T.~A.~Rijken, K.~Metzger, C.~Dullemond, G.~Rupp, and
J.~E.~Ribeiro,
Z.\ Phys.\ {\bf C30}, 615 (1986).

\bibitem{BRRD83}                 
E.~van~Beveren, G.~Rupp, T.~A.~Rijken, and C.~Dullemond,
Phys.\ Rev.\ {\bf D27}, 1527 (1983).

\bibitem{Numerical82}
C.~Dullemond, G.~Rupp, T.A.~Rijken, and E.~van~Beveren,
Comp.\ Phys.\ Comm.\ {\bf 27}, 377 (1982).

\bibitem{BDR80}            
E.~van Beveren, C.~Dullemond, and G.~Rupp,
Phys.\ Rev.\ {\bf D21}, 772 (1980);
Erratum-ibid.\ {\bf D22}, 787 (1980).

\bibitem{Bev84}                 
E.~van~Beveren,
Z.\ Phys.\ {\bf C21}, 291 (1984).

\bibitem{BR99b}                 
E.~van~Beveren and G.~Rupp,
Phys.\ Lett.\ {\bf B454}, 165 (1999)
[hep-ph/9902301].

\bibitem{Abe99}
K.~Abe {\it et al.}  [SLD Collaboration],
Phys.\ Rev.\ {\bf D59}, 012002 (1999)
[hep-ex/9805023].

\bibitem{Bev83}                 
E.~van~Beveren,
Zeit.\ Phys.\ {\bf C17}, 135 (1983).

\bibitem{BR99a}                 
E.~van~Beveren and G.~Rupp,
Eur.\ Phys.\ J.\ {\bf C11}, 717 (1999)
[hep-ph/9806248].

\bibitem{Papa01}
Joannis~Papavassiliou,
hep-ph/0102149.

\bibitem{Kota00}
V.K.B.~Kota and R.~Sahu,
nucl-th/0006079.

\bibitem{Work99}
Ron~Workman,
Phys.\ Rev.\ {\bf C59}, 3441 (1999)
[nucl-th/9811056];

Ibid.,
nucl-th/0104028.

\bibitem{Mosh91}
M.~Moshinsky,
Proceedings of the International Conference on Hadron Spectroscopy,
{\it Hadron 91}, College Park, MD, 12--16 Aug.\ 1991, p.\ 342--355.

\bibitem{Dong98}
Yu-Bing Dong,
Phys.\ Lett.\ {\bf B418}, 355 (1998).

\bibitem{Fang88}
Z.Y.~Fang, G.~Lopez~Castro, and J.~Pestieau,
Nuovo Cim.\ {\bf A100}, 155 (1988).

\bibitem{Weinberg82}
V.M.~Weinberg,
ITEP-160-1982, (Moscow, ITEP 1982).

\bibitem{Orlo84}
Marius~Orlowski,
Nuovo Cim.\ {\bf A80}, 89 (1984).

\bibitem{Est78}
P.~Estabrooks, R.~K.~Carnegie, A.~D.~Martin, W.~M.~Dunwoodie,
T.~A.~Lasinski, and D.~W.~Leith,
Nucl.\ Phys.\ {\bf B133}, 490 (1978).

\bibitem{Dum83}
O.~Dumbrajs, R.~Koch, H.~Pilkuhn, G.~C.~Oades, H.~Behrens,
J.~J.~De Swart, and P.~Kroll,
Nucl.\ Phys.\ {\bf B216}, 277 (1983).

\bibitem{Lan77}
C.~B.~Lang,
Nuovo Cim.\ {\bf A41}, 73 (1977).

\bibitem{Joh78}
N.~Johannesson and G.~Nilsson,
Nuovo Cim.\ {\bf A43}, 376 (1978).

\bibitem{Kar80}
A.~Karabarbounis and G.~Shaw,
J.\ Phys.\ G {\bf 6} (1980) 583.

\bibitem{Ber91}
V.~Bernard, N.~Kaiser, and U.~G.~Meissner,
Phys.\ Rev.\ {\bf D43}, 2757 (1991).

\bibitem{Est79}
P. Estabrooks,
Phys.\ Rev.\ {\bf D19}, 2678 (1979).

\bibitem{Ast88}
D.~Aston {\it et al.},
Nucl.\ Phys.\ {\bf B296}, 493 (1988).

\bibitem{Mat74}
M.~J.~Matison {\it et al.},
Phys.\ Rev.\ {\bf D9}, 1872 (1974).

\bibitem{Goebel00}
Carla~G\"{o}bel, on behalf of the E791 Collaboration,
hep-ex/0012009.

\bibitem{Bla100}
Deidre~Black, Amir~H.~Fariborz, Sherif~Moussa, Salah~Nasri, and
Joseph~Schechter, 
Phys.\ Rev.\ {\bf D64}, 014031 (2001)
[hep-ph/0012278].

\bibitem{Bla200}
Deirdre Black, Amir H. Fariborz, and Joseph Schechter,
Proceedings of the YITP Workshop on Possible Existence of the Sigma Meson and
its Implications to Hadron Physics, {\it Sigma Meson 2000}, Kyoto, Japan, 12-14
June 2000 [hep-ph/0008246];

Ibid.,
Proceedings of the International Workshop on Hadron Physics:
{\it Effective Theories of Low Energy QCD}, Coimbra, Portugal, 10--15 Sept.\
1999, AIP Conference Proceedings {\bf ?}, p.\ 290--299 (2000) [hep-ph/9911387];

Ibid.,
Phys.\ Rev.\ {\bf D61}, 074001 (2000)
[hep-ph/9907516].

\bibitem{Sca00}
M.D.~Scadron,
Proceedings of the YITP Workshop on Possible Existence of the Sigma Meson and
its Implications to Hadron Physics, {\it Sigma Meson 2000}, Kyoto, Japan, 12-14
June 2000 [hep-ph/0007184].

\bibitem{Jamin00}
Matthias~Jamin, Jose~Antonio~Oller, and Antonio~Pich,
Nucl.\ Phys.\ {\bf B587}, 331 (2000)
[hep-ph/0006045].

\bibitem{Ishida00}
Shin Ishida, Muneyuki Ishida, and Tomohito Maeda,
Prog.\ Theor.\ Phys.\ {\bf 104}, 785 (2000) [hep-ph/0005190]. 

\bibitem{Bab00}
L. Babukhadia, Ya.A. Berdnikov, A.N. Ivanov, and M.D. Scadron,
Phys.\ Rev.\ {\bf D62}, 037901 (2000)
[hep-ph/9911284].

\bibitem{Mar99}
V.E. Markushin and M.P. Locher,
Proceedings of the Workshop on Hadron Spectroscopy (WHS 99), Rome, Italy,
8-12 March 1999, {\it Frascati 1999, Hadron Spectroscopy}, p.\ 229--236 (1999)
[hep-ph/9906249].

\bibitem{Nap99}
J.L. Lucio Martinez and Mendivil Napsuciale,
description,''
Phys.\ Lett.\ {\bf B454}, 365 (1999)
[hep-ph/9903234].

\bibitem{Ishida99}
Muneyuki Ishida,
Prog.\ Theor.\ Phys.\ {\bf 101}, 661 (1999) [hep-ph/9902260]. 

\bibitem{Far99}
Amir H. Fariborz and Joseph Schechter,
Phys.\ Rev.\ {\bf D60}, 034002 (1999)
[hep-ph/9902238].

\bibitem{Bla98}
Deirdre Black, Amir H. Fariborz, Francesco Sannino, and Joseph Schechter,
Phys.\ Rev.\ {\bf D59}, 074026 (1999)
[hep-ph/9808415];

Ibid.,
Phys.\ Rev.\ {\bf D58}, 054012 (1998)
[hep-ph/9804273].

\bibitem{Oller99}
J.~A.~Oller and E.~Oset,
Phys.\ Rev.\ {\bf D60}, 074023 (1999)
[hep-ph/9809337].

\bibitem{Oller98}
J.~A.~Oller, E.~Oset, and J.~R.~Pelaez,
Phys.\ Rev.\ {\bf D59}, 074001 (1999)
[Erratum-ibid.\ {\bf D60}, 099906 (1999)]
[hep-ph/9804209];

Ibid.,
Phys.\ Rev.\ Lett.\  {\bf 80}, 3452 (1998)
[hep-ph/9803242].

\bibitem{Ishida97}
Shin Ishida, Muneyuki Ishida, Taku Ishida, Kunio Takamatsu, and Tsuneaki Tsuru,
Prog.\ Theor.\ Phys.\ {\bf 98}, 621 (1997) [hep-ph/9705437]. 

\bibitem{Ani97}
V.~V.~Anisovich,
Phys.\ Usp.\  {\bf 41}, 419 (1998)
[Usp.\ Fiz.\ Nauk {\bf 168}, 481 (1998)]
[hep-ph/9712504].

\bibitem{Koll00}
M.~Koll, R.~Ricken, D.~Merten, B.~C.~Metsch, and H.~R.~Petry,
Eur.\ Phys.\ J.\ {\bf A9}, 73 (2000)
[hep-ph/0008220].

\bibitem{DHM76}
R.~F.~Dashen, J.~B.~Healy, and I.~J.~Muzinich,
Phys.\ Rev.\ {\bf D14}, 2773 (1976);

Ibid.,
Ann.\ Phys.\  {\bf 102}, 1 (1976).

\bibitem{JW91}
J.~Weinstein,
Nucl.\ Phys.\ Proc.\ Suppl.\  {\bf 21}, 207 (1991);

Ibid.,
UTK-89-7,
Invited talk given at {\it Hadron '89}, Int.\ Conf.\ on Hadron Spectroscopy,
Ajaccio, France, Sept.\ 23--27, 1989;

Ibid.,
UTK-HEP-TH-88-49,
Presented at BNL Workshop on Glueballs, Hybrids, and Exotic Hadrons,
Upton, N.Y., Aug.\ 29 -- Sept.\ 1, 1988.

\bibitem{JWIs91}
J.~Weinstein and N.~Isgur,
Phys.\ Rev.\ {\bf D43}, 95 (1991);

Ibid.,
Phys.\ Rev.\ {\bf D41}, 2236 (1990).
\end{thebibliography}
\end{document}